# Evolution of Two Membrane Protein Sequences and Functions

J. C. Phillips

Dept. of Physics and Astronomy

Rutgers University, Piscataway, N. J., 08854

## Abstract

TRPC(3,6) are two ~ 930 amino acid membrane proteins that form calcium permeant cation channels. Here we examine the differences between mammals and oviparous species. Our method is based on the concept of evolution towards criticality, a general concept we have previously applied to many proteins, especially in describing the evolution of pandemic sequences through natural selection.

1.      Introduction

Here we discuss the small evolutionary differences between two membrane proteins, TRPC(3,6), that form calcium permeant cation channels. It has been estimated that membrane proteins are coded for by approximately 30% of the human genome [1]. While organic chemists have often said that "structure is function", the connection between structure and function is not easily established in membrane proteins, which cannot be studied in their native lipid environment [1]. Specifically, care must be taken even in the first step, which can be identifying transmembrane segments. Here we continue our small scale discussion of evolutionary differences, begun earlier with Piezo(1,2) [2] and MIP [3] proteins. Although we discuss only a few proteins, our method is very detailed. It has quantified the connection between sequence mutations of the Coronavirus (CoV) spike and its evolving contagiousness with high accuracy [4-7]. The method is based on a conjectured connection between living systems and evolution by natural selection towards a thermodynamic critical point, which has been popularized by physicists [8,9].

TRPC(3,6) are ~ 930 amino acid proteins, with six transmembrane segments, compared to MIP proteins, also with six transmembrane segments, and only 263 amino acids. Here we again compare humans with chickens, and find some differences attributable to chicken egg laying (oviparous), previously identified in MIP proteins [3]. We also propose small changes in transmembrane segment assignments, compared to those suggested by Uniprot.

2.    Method

Our earlier analysis relied on Ψ(aa,W) hydropathic profiles, where Ψ(aa) measures the hydropathicity of each amino acid. Small changes in protein shapes are often driven by waves in water films. These water waves have been averaged linearly over sliding windows of width W. (Data processing using sliding window algorithms is a general smoothing and sorting technique discussed online.) A natural choice for W in transmembrane (TM) proteins is 21, as used by Uniprot in listing TM segments of Piezo1 and 2. Wave motion has been an essential part of physics for centuries, but it is little used in molecular biology. It was extremely useful in analyzing the evolution of Spike sequences and connecting them to CoV contagiousness [4-7].

3.    Results

The hydropathic profiles with the KD scale (used in [2]) of human and chicken TRPC3 are shown in Fig. 1. The hydrophobic extrema (centers of transmembrane segments) identified in Uniprot Q13507 (human) are numbered 1-6. They are in overall good agreement with our hydrophobic extrema. The results for the six edges of TRPC6 (Uniprot Q9Y210), shown in Fig. 2, are similar, with one exception. We find a strong extremum near 450, labelled !, and would prefer to ignore the much weaker edge 2 of Uniprot. The figures also contain profiles for chicken TRPC3 (Uniprot F1NJJ8) and chicken TRPC6 (Uniprot A0A3Q3A7I9).

4.    Discussion

According to Uniprot Q13507, TRPC3 forms a receptor-activated non-selective calcium permeant cation channel. Looking at Fig. 1, near the N terminal we see a wedge-shaped structure with two arrows labelled ?. This structure appears suitable for opening and closing the cation channel, depending on its binding to the receptor. This happens while the transmembrane region1-6 is anchored to the membrane.

At the N-terminal edge there is a narrow strongly hydrophilic edge in humans, and an even stronger hydrophilic edge in chickens. In chickens this is caused by the long E-Q–rich sequence 19QQQQQQQQQQQQQEEEEEEEAEEEERQRRRR47, which is wedged between stiff prolines 47PP48 and 58PPPPP62. The prolines bolster the long hydrophilic sequence, which may guide contact with receptors. The weaker human edge is associated with a less hydrophilic, but still E-rich sequence 20EEEEDEGEDEGAEPQRRRR38 and bolstered by only 17PAP19 and 63PPP65. Another E-Q–rich sequence is seen in TRPC3 of American chameleons (Uniprot G1KME8).

Similarly long E-Q–rich sequences at deep hydrophilic edges were noted in human Piezo1 and Piezo2, while in TRPC3 they are seen to be longer and deeper in chicken than in human. Acute cerebellar ataxia is sudden, uncoordinated muscle movement due to disease or injury to the cerebellum. This is the area in the brain that controls muscle movement. Two mutations, R762H R847H near the C terminal (see Fig. 1), cause cerebellar ataxia [10]. Thus both Piezo1,2 and TRPC3 involve mechanically activated cation channels.

In TRPC6 (Fig. 2) there are nearly 50 mutations associated with progressive kidney failure (Uniprot Q9Y210). These mutations could depress channel management. We divide proteins into three groups (Hydrophilic: D, E, K, N, Q, R, S, T; Hydrophobic: C, F, I, L, M, V, W, Y; and Hydroneutral: A, G, H). Then most mutations mix philic with other philics, and phobics with other phobics. However, there are nearly twice as many philic-phobic mutations as phobic-philic. These mutations would make mutated TRPC less flexible. There are also three proline mutations (P112Q, near the N terminal; P924S, near the C terminal, which make both ends more flexible. An exceptional case is L780P. This coincides with the hydrophobic secondary extrema in Fig. 2, which becomes less hydrophobic, but is stiffened by the addition of proline.

What causes chickens and lizards to have long sequences in E and Q, much longer than in mammals? It could be the difference in chemistry at gestation, for instance, high concentrations of hen egg white in eggs [11].

Figure Captions.

Fig. 1. Hydroprofile of TRPC3. The mechanical significance of the wedge near 200 is discussed in the text. Site numbering for chicken. Small blue dots near the C terminal mark two disease mutations.

Fig.2. Hydroprofile of TRPC6. For human site nos., subtract 38.

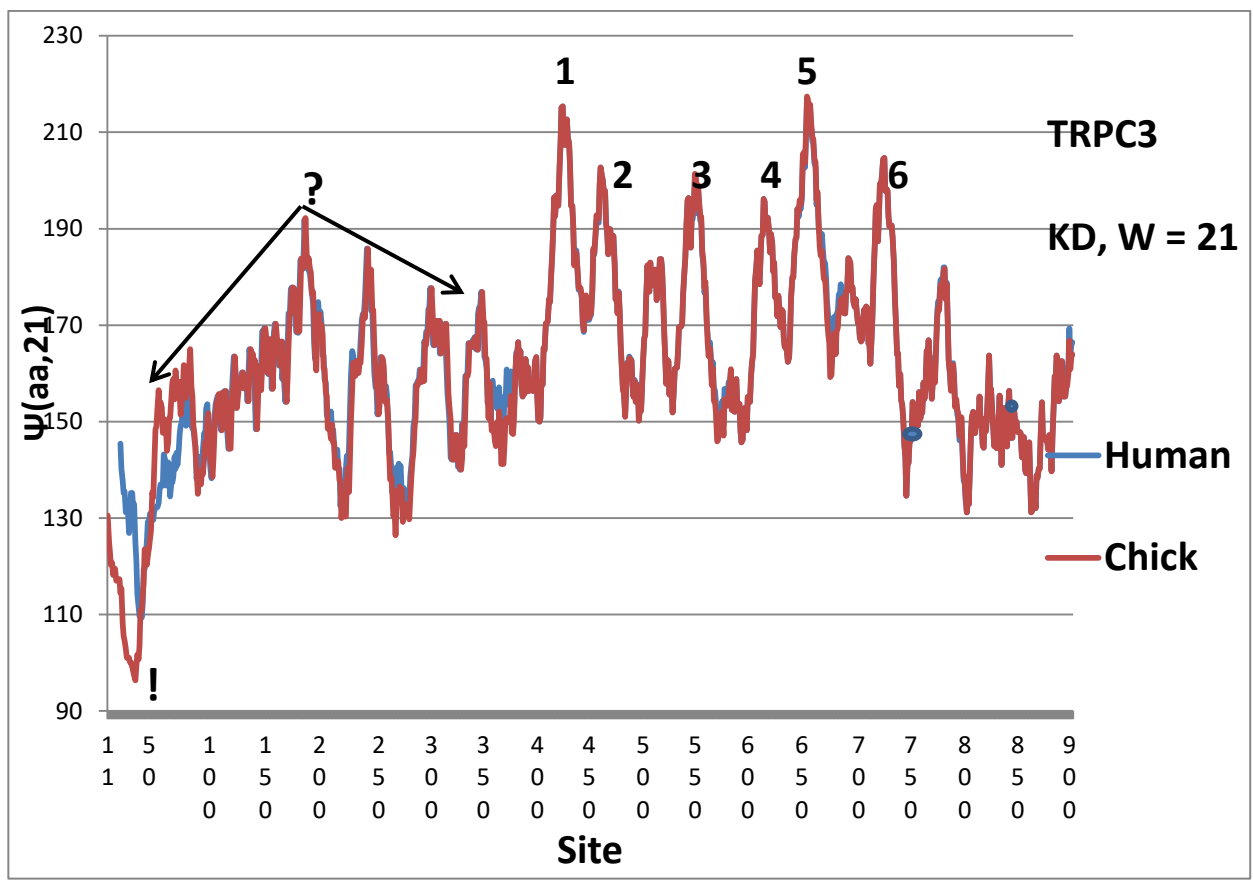

Fig. 1.

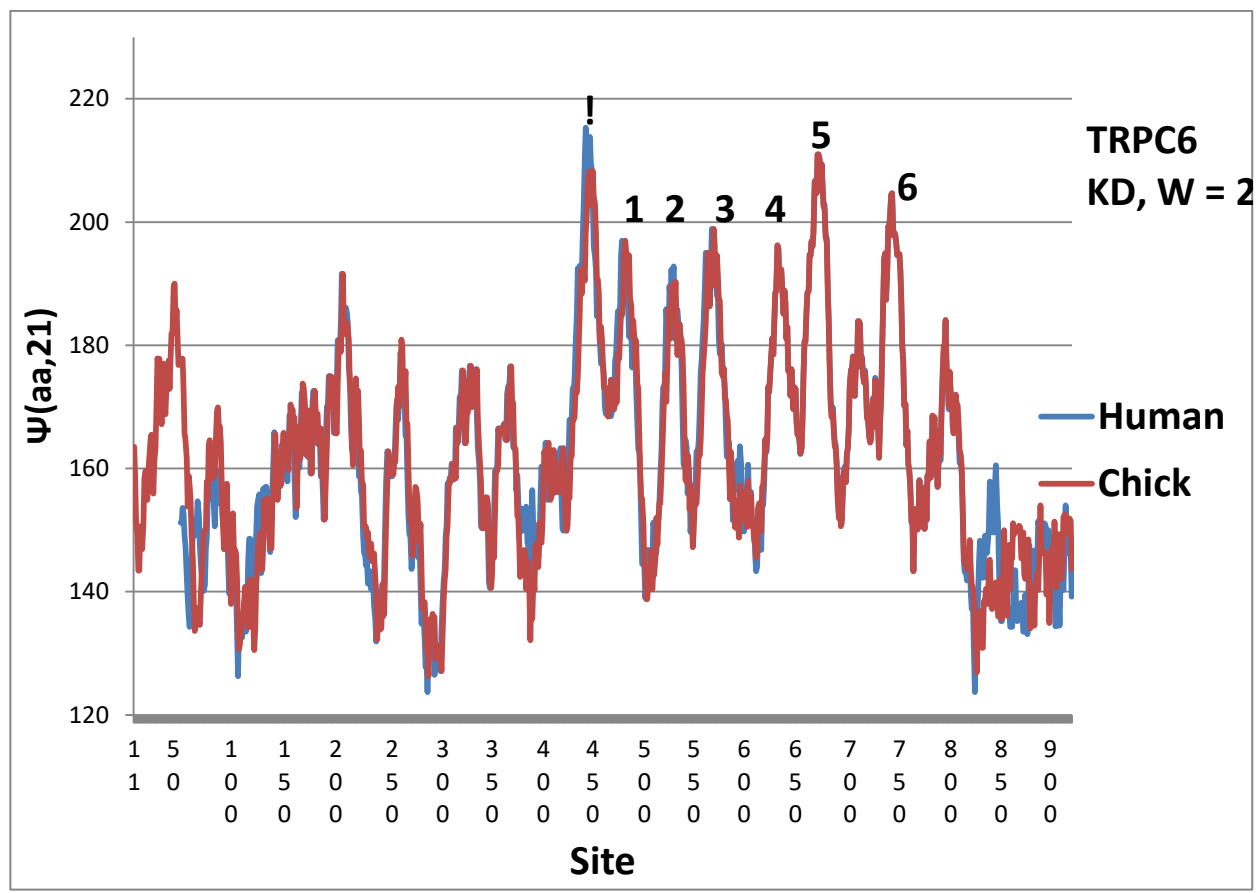

Fig. 2.